\documentclass[sigconf]{acmart}

\usepackage{hyperref}
\usepackage{url}
\usepackage{graphicx} 
\usepackage{multirow,diagbox,enumitem}
\usepackage{booktabs}
\usepackage{arydshln}
\usepackage{subfig}
\usepackage{algorithm}
\usepackage{amsthm}
\usepackage{makecell}
\usepackage{xcolor}
\usepackage{bm}
\usepackage{arydshln}
\usepackage{subcaption}

\usepackage{algorithmic}
\usepackage[most]{tcolorbox}
\usepackage{tabularx}
\newcommand{\cmark}{\ding{51}} % ✔
\newcommand{\xmark}{\ding{55}} % ✘
\usepackage{pifont}
\AtBeginDocument{%
  }

\copyrightyear{2026}
\acmYear{2026}
\setcopyright{cc}
\setcctype{by}
\acmConference[WWW '26] {Proceedings of the ACM Web Conference 2026}{April 13--17, 2026}{Dubai, United Arab Emirates.}
\acmBooktitle{Proceedings of the ACM Web Conference 2026 (WWW '26), April 13--17, 2026, Dubai, United Arab Emirates}
\acmISBN{979-8-4007-2307-0/2026/04}
\acmDOI{10.1145/3774904.3792555}
\theoremstyle{definition}

\theoremstyle{plain}

\newcommand{\vpara}[1]{\vspace{0.05in}\noindent\textbf{#1 }}
\newcommand{\ours}{PolicySim}

\newcommand{\hide}[1]{}

\begin{document}

\title{PolicySim: An LLM-Based Agent Social Simulation Sandbox for Proactive Policy Optimization}

\author{Renhong Huang}
\affiliation{
  \institution{Zhejiang University}
  \city{Hangzhou}
  \country{China}
  }
\email{renh2@zju.edu.cn}

\author{Ning Tang}
\affiliation{
  \institution{Fudan University}
  \city{Shanghai}
  \country{China}
  }
\email{ningtang24@m.fudan.edu.cn}

\author{Jiarong Xu}
\affiliation{%
  \institution{Fudan University}
  \city{Shanghai}
  \country{China}
}
\email{jiarongxu@fudan.edu.cn}
\authornote{Corresponding author.}

\author{Yuxuan Cao}
\affiliation{
  \institution{HKUST}
  \city{Hong Kong}
  \country{China}
  }
\email{ycaoce@connect.ust.hk}

\author{Qingqian Tu}
\affiliation{%
  \institution{University of Nottingham}
  \city{Nottingham}
  \country{United Kingdom}
}
\email{psxqt2@nottingham.ac.uk}

\author{Sheng Guo}
\affiliation{
  \institution{MyBank, AntGroup}
  \city{Hangzhou}
  \country{China}
  }
\email{guosheng.guosheng@mybank.cn}

\author{Bo Zheng}
\affiliation{
  \institution{MyBank, AntGroup}
  \city{Hangzhou}
  \country{China}
  }
\email{guangyuan@mybank.cn}

\author{Huiyuan Liu}
\affiliation{
  \institution{PowerChina Huadong Engineering Corporation Limited}
  \city{Hangzhou}
  \country{China}
  }
\email{liu_hy8@hdec.com}

\author{Yang Yang}
\affiliation{%
  \institution{Zhejiang University}
  \city{Hangzhou}
  \country{China}
  }
\email{yangya@zju.edu.cn}

% \renewcommand{\shortauthors}{Renhong Huang, Ning Tang, Jiarong Xu, \& Yang Yang}
\renewcommand{\shortauthors}{Renhong Huang et al.}



% \keywords{Do, Not, Use, This, Code, Put, the, Correct, Terms, for,




\begin{abstract}
Social platforms serve as central hubs for information exchange, where user behaviors and platform interventions jointly shape opinions. However, intervention policies like recommendation and content filtering, can unintentionally amplify echo chambers and polarization, posing significant societal risks. Proactively evaluating the impact of such policies is therefore crucial. Existing approaches primarily rely on reactive online A/B testing, where risks are identified only after deployment, making risk identification delayed and costly. LLM-based social simulations offer a promising pre-deployment alternative, but current methods fall short in realistically modeling platform interventions and incorporating feedback from the platform.  Bridging these gaps is essential for building actionable frameworks to assess and optimize platform policies. To this end, we propose PolicySim, an LLM-based social simulation sandbox for the proactive assessment and optimization of intervention policies. PolicySim models the bidirectional dynamics between user behavior and platform interventions through two key components: (1) a user agent module refined via supervised fine-tuning (SFT) and direct preference optimization (DPO) to achieve platform-specific behavioral realism; and (2) an adaptive intervention module that employs a contextual bandit with message passing to capture dynamic network structures. Experiments show that PolicySim can accurately simulate platform ecosystems at both micro and macro levels and support effective intervention policy.
% Additionally, we have extensively demonstrated the effectiveness of proactive strategy for adaptive intervention policy.
\vspace{-0.1in}
\end{abstract}

\maketitle

\vspace{-0.1in}
\section{Introduction}
\begin{figure}[t]     
    \centering
    {\includegraphics[width=1.0\columnwidth]{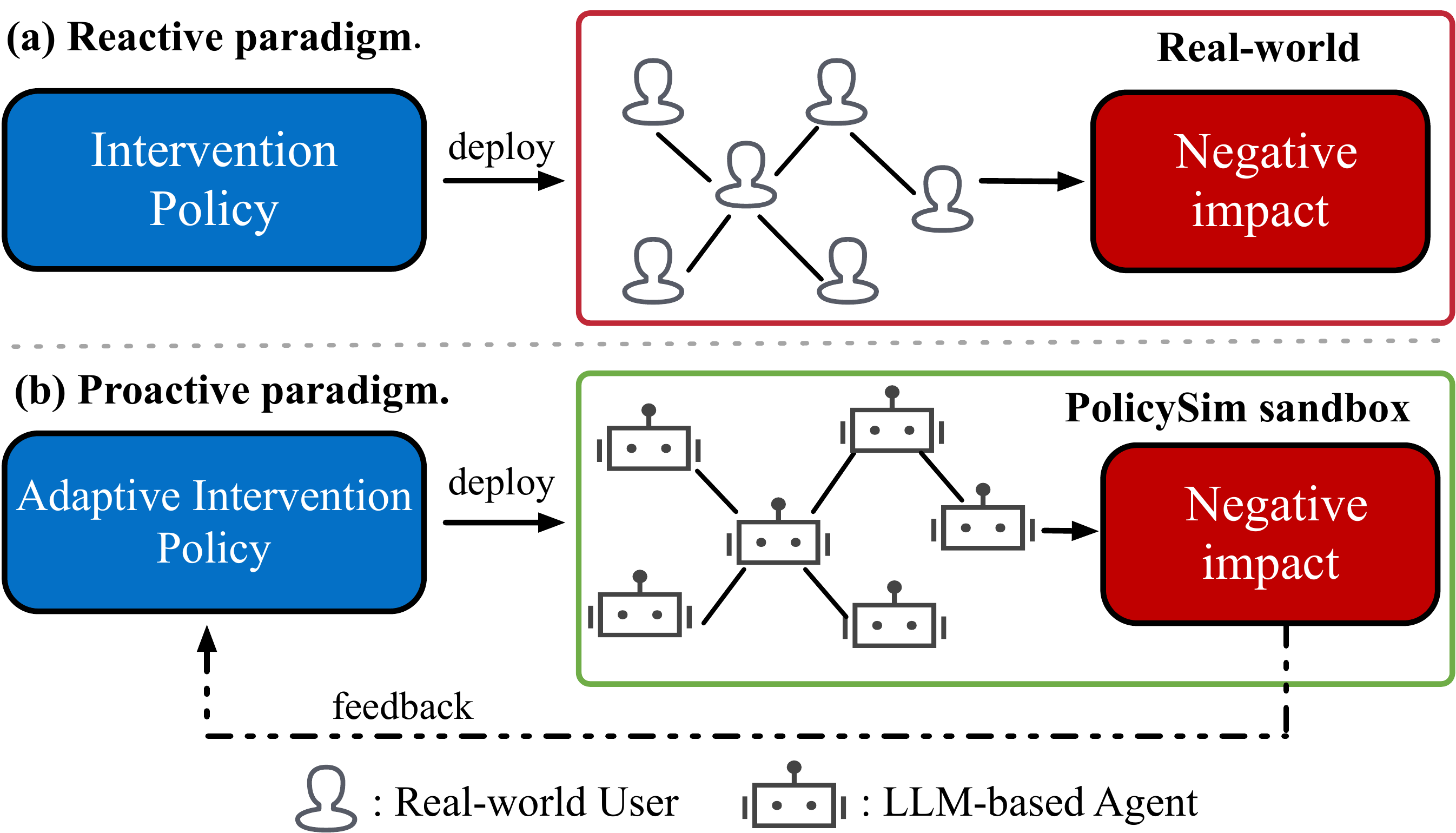}}
    % \vspace{-}
    \caption{Compared to A/B testing, which \emph{reactively} assesses interventions by deploying and learning only after outcomes are observed, \ours~\emph{proactively} assesses and optimizes intervention policies prior to deployment via feedback.
    % that closely mirrors real-world dynamics.
    }\label{fig:intro}
    \vspace{-0.3in}
\end{figure}
\noindent
In today's digital era, social platforms have become the core infrastructure for social interaction and information exchange~\cite{kietzmann2011social,kaplan2010users,van2013culture}. 
Platform intervention policies fundamentally shape users’ interactive behaviors including information consumption, content sharing, and social engagement, through intervention mechanisms such as recommender systems~\cite{ko2022survey, zhao2024recommender,xu2025user, yuan2025tree}, content filtering~\cite{rodgers2024enhancing}, and exposure control~\cite{fletcher2018people}. Collectively, these intervention mechanisms shape users’ online experiences and exert a profound influence on opinion formation and decision-making processes~\cite{gao2023s,cinelli2021echo,fletcher2018people}.

However, inappropriate platform intervention policies can trigger unintended and even deleterious consequences, as social platforms’ inherent tendency to amplify information can magnify the negative effects of such policies beyond their initial expectations~\cite{ko2022survey,zhao2024recommender,rodgers2024enhancing}.  For instance, numerous studies have demonstrated that recommender systems can foster the emergence of echo chambers and filter bubbles~\cite{cinelli2021echo,bozdag2015breaking}, thereby suppressing cross-viewpoint dialogue and reducing stance diversity~\cite{barbera2015birds}. At the same time, platform interventions that prioritize user engagement may increase polarization and conflict, undermining public trust and attracting regulatory scrutiny~\cite{persily2020social}.
Consequently, a key research challenge is to \emph{assess the potential social impacts of intervention policies in a proactive and systematic manner prior to deployment}.

Existing evaluation methods primarily rely on the A/B testing to assess intervention policies~\cite{gilotte2018offline}, requiring direct deployment in real user environments to collect behavioral feedback, as illustrated in Figure~\ref{fig:intro}(a). However, this paradigm suffers from several limitations: (i) evaluation is reactive rather than proactive; (ii) feedback loops are often delayed and therefore cannot keep pace with rapid platform dynamics; and (iii) directly testing in real-world environments may introduce uncontrollable and potentially irreversible harmful consequences. Thus, reliance on A/B testing alone is insufficient for the prospective evaluation of intervention policies, highlighting the need for proactive, risk-aware alternatives.

Recent advances in large language model (LLM)-based simulation, enabled by their strong generative and reasoning capabilities, offer a promising direction. Prior works include traditional agent-based models~\cite{jackson2017agent, schelling2006micromotives} and LLM-based agent simulation frameworks~\cite{chen2023agentverse, liu2024lmagent, mou2024unveiling, yang2024oasisopenagentsocial}. Both types of models have been used to study complex social phenomena, including opinion dynamics~\cite{chuang2023opinion}, economic systems~\cite{li2023econagent}, and social norm alignment~\cite{ren2024emergence}. Nevertheless, several challenges remain unresolved in current works. (i) Most simulations do not explicitly model platform intervention policies, making it difficult to accurately capture their effects. (ii) Their agent design often relies heavily on prompt engineering rather than realistic modeling of social media behavior, which limits the credibility of simulated outcomes. (iii) Existing frameworks lack principled mechanisms for leveraging simulation feedback to optimize real-world intervention policies, limiting the use of simulation outcomes for iterative improvement of real-world policies.

To bridge these gaps, we present \ours, a LLM-based multi-agent social simulation sandbox for the proactive assessment and optimization of intervention policies. The framework captures the bidirectional dynamics between intervention strategies and ecosystem evolution, as well as the influence of interventions on user behavior patterns. To achieve this, the framework introduces a user agent module and an intervention policy module. Specifically, the user agent module is introduced as a novel paradigm for training social agents that integrates supervised fine-tuning (SFT) with direct preference optimization (DPO). This unified approach jointly enhances the behavioral faithfulness of agents to platform-specific user data and improves the distinctiveness of user intent representations. Building upon the simulation, the intervention policy module is introduced to collect rewards from the sandbox and adaptively optimize  intervention policies. This module employs a contextual bandit algorithm that balances exploration and exploitation while leveraging message passing to capture dynamic user networks. We incorporate multiple intervention policies, such as \emph{recommender systems} and \emph{exposure control}), deployed on X and Weibo platforms. 
% Finally, we validate the effectiveness of \ours~and evaluate the intervention module with respect to predefined objectives. Comprehensive experimental results demonstrate that \ours~achieves highly effective policy optimization in complex social ecosystems. 

% \vspace{-0.3in}
\begin{table}[t]
    \centering
    \setlength{\tabcolsep}{1.5pt}
     \renewcommand{\arraystretch}{1.05}
    \resizebox{0.9\columnwidth}{!}{ 
    \centering
    \renewcommand{\arraystretch}{1.3}
    \begin{tabular}{c|ccccc}
    \toprule[1pt]
        & Scale & Relation & IP. & AI. &Env.\\ \hline
        \ours&1000 &\cmark &\cmark &\cmark & X \& Weibo \\ \hline
        Oasis~\cite{yang2024oasisopenagentsocial}  &1M &\cmark &\cmark &\xmark & X \& Reddit \\
        Agent4rec~\cite{zhang2024generative}  &1000 &\cmark &\cmark &\xmark &Movie Rec\\
        HiSim~\cite{mou-etal-2024-unveiling}&700& \xmark &\xmark &\xmark &X\\
        Stopia~\cite{zhou2025sotopia} &2& \xmark &\xmark &\xmark & -\\ \bottomrule
    \end{tabular} }
    \caption{Comparison of our system with recent social simulation frameworks. Scale: number of LLM agents; Relation: whether follow/follower links evolve; IP/AI: presence of intervention policy and adaptive interventions; Env.: underlying platform (``–'' means unspecified).}
    \vspace{-0.4in}
\end{table}

Our contributions are summarized as follows:

% % \vspace{-0.05in}
\begin{itemize}[leftmargin=10pt]
\item We propose \ours, a LLM-based multi-agent social simulation sandbox for the proactive assessment and optimization of intervention policies. 
\item To enhance LLM-based agent simulation fidelity, we propose, for the first time, a unified paradigm for training social agents that combines
% a social-agent training paradigm combining 
SFT and DPO, which ensures behavioral alignment with platform data while capturing diverse user intents.
% and diverse of user intents.
\item To enable adaptive optimization of 
intervention policy, we employ
% using 
a contextual bandit framework that balances exploration and exploitation, augmented with message passing to capture dynamic network structures and information flows.
\item Extensive experiments across multiple datasets verify the realism of agent behaviors and the effectiveness of intervention optimization, showing that \ours~enables scalable and proactive assessment of intervention policies.
\end{itemize}
% \vspace{-0.05in}

The paper is organized as follows. \S\ref{sec:framework} presents the social simulation sandbox, comprising user-agent and intervention-policy modules. Building on this sandbox, \S\ref{sec:optimization} introduces  adaptive intervention policy based on a bandit algorithm for automated feedback and optimization. Finally, \S\ref{sec:exp} evaluates the effectiveness of our framework through simulation and intervention experiments.

\vspace{-0.1in}
\section{Simulation Sandbox Framework}\label{sec:framework}
\begin{figure*}[!t]
    \centering
    \includegraphics[width=0.9\textwidth]{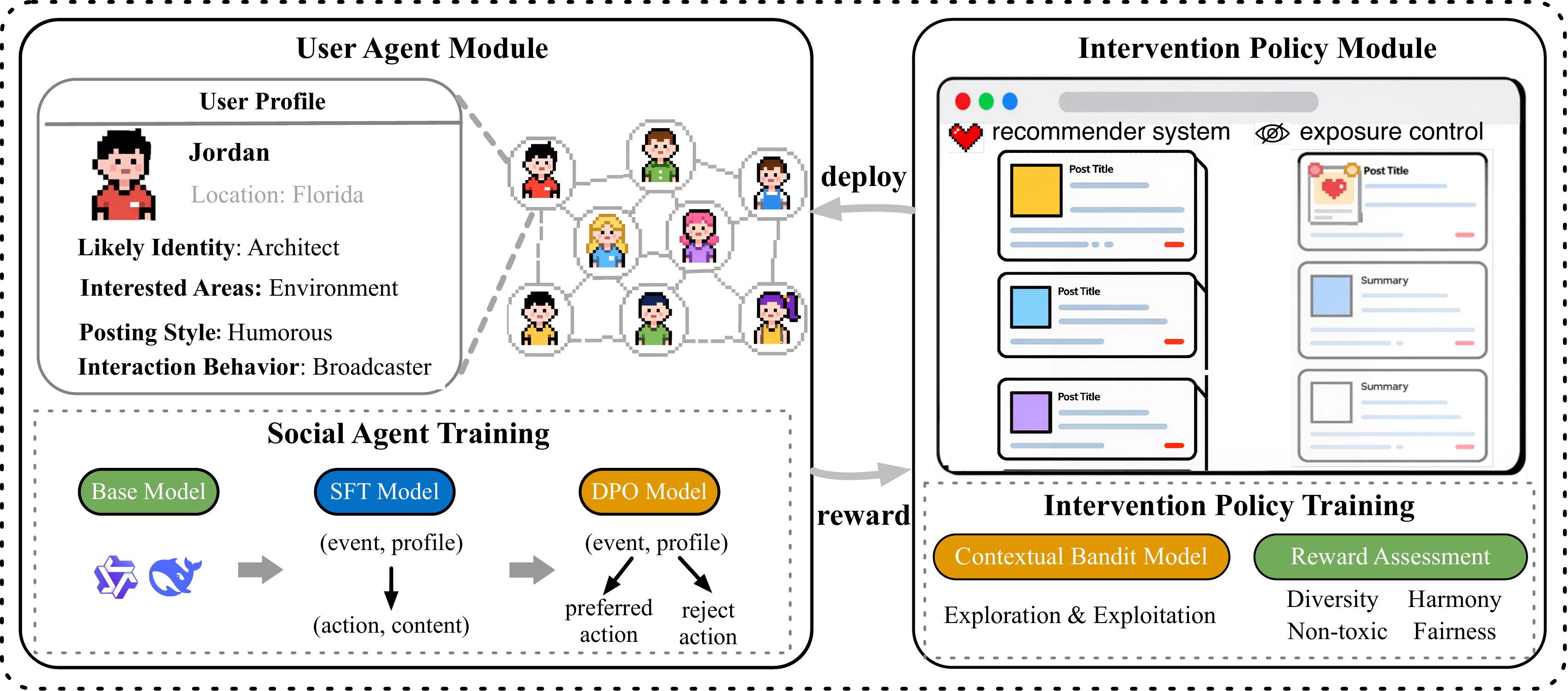}
    \caption{The architecture of \ours~is composed of two main modules: the User Agent Module and the Intervention Policy Module. The User Agent Module contains detailed components including user profiles, memory, user relations, and behavioral models. The bottom-left panel illustrates how we train agent tailored for social media environments to capture realistic user behaviors. The Intervention Policy Module instantiates typical platform mechanisms such as recommender systems and exposure control. By simulating intervention policies within sandbox framework, the Target Reward Assessment component evaluates their performance and provides feedback, which are further utilized to adaptively optimize intervention policy.}\label{fig:framework}
\end{figure*}

In this section, we first present the framework of \ours, as illustrated in Figure~\ref{fig:framework}. We primarily follow the architecture of HiSim~\cite{mou-etal-2024-unveiling} to help us build the framework. The entire framework consists of the user agent module (see \S \ref{subsec:user-agent}) and the intervention policy module (see \S \ref{subsec:intervention-policy}). Each agent represents an LLM-powered user with authentic profile, accumulates memories across simulation rounds, and can simulate the interaction between intervention policies and social ecosystem evolution.

\subsection{User Agent Module} \label{subsec:user-agent} 
To enable LLMs to simulate social user behaviors, we equipped agents with specialized modules including user profile, user behavior, memory, training and planning. Based on these designs, agents are empowered to emulate realistic user decision-making processes.

\subsubsection{User Profile.} User profiles are crucial for capturing interests, behaviors, and preferences, supporting tasks such as personalized recommendation, behavior prediction, and social agent modeling~\cite{zhang2024generative, wu2024personalized, park2022social}. As explicit profiles are mostly unavailable, we construct them from post content and user metadata, extracting four high-level attributes for agent prompting:
\begin{itemize}[leftmargin=10pt]
\item Likely identity: Infers social or professional roles, enabling domain-specific terminology and perspectives.
\item Interested areas: Identifies user interests from posts, hashtags, and retweeted accounts (e.g., a user engaging with $\#$ ClimateEmergency is labeled with an environment-related interest).
\item Posting style. Linguistic traits inferred from posts, emojis, slang, and sentiment.
\item Interaction behavior. Quantifies engagement via post/retweet ratios, reply frequency, and follower/following ratios, categorizing users by interaction roles.
\end{itemize}
Let $U$ be the set of users. For each $u_i \in U$, we denote their profile by $\phi(u_i)$. Details on the extraction of the four high-level attributes and the social media metadata format are provided in Appendix~\ref{app:exp-setup}.

\subsubsection{User Behavior.} \label{subsubsec:behavior}
During each simulation round, agents undergo behaviors calling phase, interacting with the environment and executing various behaviors. Our agent architecture extends conventional social interactions~\cite{mou2024unveiling} (tweet, retweet, reply, like, dislike, do nothing) by adding two relationship actions, \emph{follow} and \emph{unfollow}, enabling the network topology to evolve dynamically in real time according to agents’ content preferences and behaviors. Unlike prior work~\cite{mou2024unveiling,mi2025mf} executing a single action per round, we implement \emph{Multiple Behavior Selection} in the behavior calling phase, capturing diverse engagement patterns and enhancing interaction variety to better reflect real-world social media dynamics. Detailed definitions of behaviors and the prompts used for behavior generation are provided in Appendix~\ref{app:exp-setup}.

\subsubsection{Relation.} The social media network is typically modeled as dynamic graph $G_t = (V, E_t)$ at simulation round $t$, where $V$ is the set of users and $E_t$ contains directed edges representing follow relationships. An edge $e_{ij} \in E_t$ indicates that agent $u_i$ follows or interacts with agent $u_j$ at round $t$. The network evolves over time, with edges updated dynamically across rounds due to \emph{follow} and \emph{unfollow} actions, yielding a time-varying topology $\{G_1, \dots, G_T\}$.

\subsubsection{User Stance.} To model group opinion dynamics on social media, we quantify user agents’ stances toward events using LLMs, which capture implicit attitudes and subtle textual patterns, and have shown high accuracy in stance detection~\cite{lan2024stance, cruickshank2023use}.

Specifically, a user’s stance at simulation round $t$ is discretely classified as ${-1,0,1}$ for negative, neutral, and positive, respectively. For user $u_i$, the discrete stance score is inferred as:
\begin{equation}
\bar{s}^t(u_i) = f_{\tau_1}\big(p(u_i), \mathrm{Action}(u_i)[:t]\big) \in \{-1, 0, 1\},
\end{equation}
where $f_{\tau_1}$ denotes the LLM conditioned on prompt $\tau_1$ and the user’s history. Further, to reduce noise from LLM hallucinations~\cite{ji2023towards}, we apply an exponential moving average:
\begin{equation}\label{eq:stance_update}
s^t(u_i) = \alpha s^{t-1}(u_i) + (1 - \alpha) \bar{s}^t(u_i),
\end{equation}
with smoothing coefficient $\alpha \in (0,1)$ and initialization $s^0(u_i) = \bar{s}^0(u_i)$. A larger $\alpha$ emphasizes past stances in the current score.

\subsubsection{Memory.} Memory encodes, stores, and retrieves information to influence future actions~\cite{sherwood2004human,atkinson1968human}. We model user memory with short-term and long-term components to capture dynamic information retention and retrieval~\cite{norris2017short}.

\vpara{Short-term memory.} This component handles temporary storage and fast processing. Humans typically retain only the most salient content, varying across messages and contexts. Therefore, for user $u_i$, the $k$-th short-term memory $m_k$ is generated as:
\begin{equation}
m_k = f_{\tau_2}(c, \phi(u_i), m_{k-1}),
\end{equation}
where $c$ is post content and $f_{\tau_2}$ directs the LLM via prompt $\tau_2$. Memories are stored in \emph{memory pool} $\{m_k, \mathrm{emb}(m_k), t\}$, including content, embedding by embedding model, and simulation round.

\vpara{Long-term memory.} Long-term memory retains information of lasting significance, such as past experiences or high-level insights. Retrieval samples memories from the pool, prioritizing those with high semantic relevance and accounting for temporal decay:
\begin{equation}
\text{Pr}(m_k) \propto e^{-\lambda \Delta_t} \text{Sim}(\mathrm{emb}(c), \mathrm{emb}(m_k)),
\end{equation}
where $\lambda>0$ is a decay rate, $\Delta_t$ is the elapsed time, and $\text{Sim}(\cdot,\cdot)$ measures semantic similarity. Sampled short-term memory entries $\{m\}$ are integrated into long-term memory $\hat{m}_k$ via $\hat{m}_k = f_{\tau_3}(c, \phi(u_i), \{m\})$ with $f_{\tau_3}$ guiding the LLM using prompt $\tau_3$.

To reduce computational cost during memory generation process, short posts below a length threshold bypass LLM processing and are directly added to the memory pool. Full algorithmic details are provided in Appendix~\ref{app:exp-setup}.

\subsubsection{Agent Training and Planning} Existing multi-agent designs largely rely on prompt engineering. While these expert-crafted prompts can improve plausibility, they lack alignment with real-world social media data. Social media agents exhibit heterogeneous, context-sensitive behaviors with long-term dependencies. Therefore, to better capture authentic user behaviors, we train agents directly on platform data rather than relying on prompt-based heuristics, and we adopt a two-stage framework combining Supervised Fine-Tuning (SFT) and Direct Preference Optimization (DPO) to preserve both stylistic and behavioral consistency.

\vpara{SFT as cold start.} Given a collection of (event, user, action) tuples $\{(e_i, u_i, a_i)\}_{i=1}^{M}$, we first construct a dataset of instruction-action pairs $\mathcal{D}_{\text{SFT}} = \{(\bm{x}_i, \bm{y}_i)\}_{i=1}^M$. Each instruction sequence $\bm{x}_i$ concatenates the topic description and the associated user profile, i.e. $\bm{x}_i = [e_i, \phi(u_i)]$, which jointly defines the situational context and the characteristics of target user. The response $\bm{y}_i$ combines observed user action with its corresponding textual content. The policy model, parameterized by $\theta$, is first trained to minimize the negative conditional log-likelihood over the response sequence:
\begin{equation}
\label{eq:sft_loss}
\mathcal{L}_{\text{SFT}}(\theta) = -\sum_{i=1}^{N} \sum_{t=1}^{L_i} \log P\left(y_{i,t} \mid \bm{x}_i, \bm{y}_{i, <t}; \theta\right),
\end{equation}
where $L_i$ is the length of the response sequence for the $i$-th sample and $y_{i, t}$ denotes the $t$-th token in the sequence.

\vpara{RL Post Training.} With SFT as semantic grounding, we further employ Direct Preference Optimization (DPO)~\cite{rafailov2023direct}, explicitly aligning the agent's behaviours with desired social behaviors observed. We construct a preference dataset $\mathcal{D}_{\text{DPO}} = \{(\bm{x}_i, \bm{y}_i^{+}, \bm{y}_{i,j}^{-})_{j=1}^{J}\}_{i=1}^{M}$, where $\bm{x}_i$ represents the same prompt used in SFT, $\bm{y}_i^{+}$ denotes the preferred response, and $\bm{y}_i^{-}$ is a rejected alternative generated by the pretrained model under the same prompt. Specifically, multiple candidate actions for a given $(e_i, u_i)$ pair are produced by prompting the base model to generate. Among these, $J$ samples that exhibit low semantic similarity to $\bm{y}_i^{+}$ or differ in action choice are retained as negative options to guide preference learning.

Starting from a reference policy $\pi_{\text{ref}}$ initialized with the SFT model, DPO further optimizes the policy $\pi_{\theta}$ by maximizing the DPO loss $\mathcal{L}_{\text{DPO}}({\theta})$, formulated as:
\begin{equation}
- \mathbb{E}_{(\bm{x}, \bm{y}^+, \bm{y}^-)\sim\mathcal{D}_{\text{DPO}}}
    \Bigg[
    \log \sigma \Big(
    \beta \Big[
    \log \frac{\pi_{{\theta}}(\bm{y}^+ \mid \bm{x})}{\pi_{\text{ref}}(\bm{y}^+ \mid \bm{x})} 
    - 
    \log \frac{\pi_{\bm{\theta}}(\bm{y}^- \mid \bm{x})}{\pi_{\text{ref}}(\bm{y}^- \mid \bm{x})}
    \Big]
    \Big)
    \Bigg].
\end{equation}
This objective guides the policy with users' behavioral preferences, enabling the agent to align its responses with realistic social actions observed in human data. 

We also compared different training strategies as well as prompt-instructed agents, the experiment results can be found in \S~\ref{sec:exp}. Furthermore, we utilize CoT reasoning~\cite{wei2022chain} to enhance the interpretability of agent behaviors. In addition, we further provide a theoretical discussion on the rationale of using LLMs for learning compared to relying solely on ABM models as multi-agent systems, which would produce inseparable “users” rather than realistic user populations. Details are provided in Appendix~\ref{app:proofs}.

\subsection{Intervention Policy Module}\label{subsec:intervention-policy}
There are various types of intervention policies on social platforms, which can have profound impacts on user behavior, information propagation, and platform ecosystems. In this section, we introduce commonly adopted intervention policies in social media scenarios, along with the typical objectives these interventions aim to achieve.

\subsubsection{Intervention Policy}
In our work, we primarily focus on the following two intervention policies.

\vpara{Recommender System.} The recommender system acts as a control mechanism for regulating information access, thereby playing a pivotal role in shaping the dynamics of information flow within the platform~\cite{lu2015recommender}. Typically, the recommender system delivers relevant content aggregated from three primary sources~\cite{melville2010recommender}:

\begin{itemize}[leftmargin=10pt]
    \item Relational recommendation: Posts from users that an agent follows, reflecting direct social connections. A message posted at time $t$ becomes visible to followers at $t+1$.
    \item Personalized recommendation: The dominant channel in social platforms, delivering content tailored to individual preferences by prioritizing posts semantically aligned with users’ historical behavior~\cite{liang2006personalized} and profile representations.
    \item Headline recommendation: This channel provides non-personalized content such as trending topics or headline news, curated to highlight widely popular information.
\end{itemize}
By combining these channels, the recommender system regulates the information accessible to each user, shaping engagement patterns and balancing individual and global exposure.

\vpara{Exposure control mechanisms.}  Exposure control mechanisms regulate the visibility of content to specific user groups, serving as a tool to simulate moderation, content prioritization, or fairness-oriented interventions~\cite{togashi2024scalable,mansoury2021graph}. Formally, for a given user $u_i \in U$, we define its exposure probability at time step $t$ as $\exp(u_i) \in [0,1]$, representing the likelihood that a post generated by $u_j$ passes the platform’s filtering. By adjusting parameter $\exp(u_i)$ for each user, the platform can increase or suppress the exposure of particular user groups, thereby emulating interventions such as promoting underrepresented content, reducing the spread of misinformation, or mitigating echo chamber effects~\cite{conti2024revealing}.

\subsubsection{Intervention Objective.}\label{subsubsec:intervention} The objectives of intervention policies in social media are to guide platform dynamics towards desirable outcomes. Specifically, our interventions are designed to achieve the following goals separately:

\begin{itemize}[leftmargin=10pt]
\item \textbf{Promote cross-viewpoint interactions.} Encourage engagement between users with opposing stances, fostering diverse perspectives without increasing toxicity~\cite{törnberg2023simulatingsocialmediausing, mou2024unveiling}.

\item \textbf{Mitigate misinformation.} Limit the visibility and propagation of misleading content, improving information reliability.

\end{itemize}

\vspace{-0.1in}
\section{Adaptive Intervention Policy}\label{sec:optimization}
Built on the \ours~sandbox, we obtain feedback from multi-agent interactions in the simulated environment. These signals guide the adaptive optimization of the intervention policy toward target objectives. We next formalize the problem definition.

\vpara{Problem Definition:} Consider a social platform $\mathcal{S}_t = (U, G_t, R)$ at round $t$, where $U$ is the user set, $G_t$ the social network induced by user interactions, and $R$ the intervention policy. Let $\rho(\mathcal{S}_t)$ be a utility function measuring how well the platform objectives are achieved. As for adaptive intervention policy, agents interact under policy $R$, producing feedback signals that reflect changes in $\rho(\mathcal{S}_t)$. These signals are then used to adaptively update $R$ within the action space $\mathcal{A}$ to maximize the expected utility $\mathbb{E}[\rho(\mathcal{S}_t)]$.

This problem is practically motivated. For instance, when the intervention policy involves recommender systems, it typically relies on large amounts of historical data for offline training~\cite{khusro2016recommender, koren2009matrix}. A key shortcoming of such policy is their inability to optimize using interaction signals derived from multi-agent environments.

\subsection{Adaptive Interventions via RL} The adaptive intervention problem can be naturally framed as reinforcement learning task: By interacting with the environment, we can apply desired impact as a reward and use appropriate reinforcement learning algorithms to enable the agent to achieve the maximum cumulative reward in a dynamic environment.

To enable real-time adaptation to evolving user behaviors, we employ contextual multi-armed bandits, which provide a lightweight and flexible framework~\cite{ban2021ee,kassraie2022graph,li2010contextual}. Formally, intervention process is modeled over $T$ rounds. At each round $t$, the policy selects from $n$ arms, $X^t = {x_1^t, \dots, x_n^t}$, where each arm represents a candidate action from the action space $\mathcal{A}$. Pulling arm $x_i^t$ yields a reward:$r_i^t = \psi(x_i^t) + \xi_i^t,$ where $\psi$ maps the arm’s context to the reward, and $\xi_i^t$ is zero-mean noise. Following prior work~\cite{ban2021ee}, rewards are bounded in $[0,1]$.

\vpara{Action Space.} The action space depends on the type of intervention policy. For recommender systems, each arm represents a user-post pair, whereas for exposure control mechanisms, each arm corresponds to a user-probability pair. After defining the arm type, we construct a discrete action space by sampling candidate sets. For recommender systems, we sample the post set ${P}\mathrm{cand} \subseteq P_t$ from historical posts and the user set ${U}\mathrm{cand} \subseteq U$, preferring posts not previously recommended. The final candidate arms are the Cartesian product ${U}\mathrm{cand} \otimes {P}\mathrm{cand}$.

\vpara{Context Design.}  Context embedding of each arm combines information from both the user and the post. For a user $u_i$, we define the context embedding as:
\begin{equation}
X_{\mathrm{user}}(u_i) = \mathrm{emb}\Big([\,\phi(u_i) : \{m\}\,]\Big),
\end{equation}
where $\phi(u_i)$ is the user profile and $\{m\}$ represents the user’s recent memory. $X_{\mathrm{user}}$ denotes the context embedding matrix for all users.  

To capture social influence, we further propagate context embeddings across the social network, inspired by label propagation~\cite{zhu2003semi}. Let $G_t$ be the social graph with adjacency matrix $A$ and degree matrix $D$. We iteratively update user embeddings via:
\begin{equation}
X_{\mathrm{user}}^{k} = \gamma X_{\mathrm{user}}^{k-1} + (1-\gamma) D^{-1} A X_{\mathrm{user}}^{k-1},
\end{equation}
where $\gamma \in [0,1]$ balances self-information and neighbor aggregation. After $k$ iterations, each user embedding incorporates from $k$-hop neighbors. Finally, we concatenate the propagated user embedding with the post’s embedding to obtain the arm context.

\vpara{Reward Assessment.} The reward is equal to $\rho(\mathcal{S}_t)$, reflecting the objective of the adaptive intervention algorithm. Different intervention goals correspond to different reward formulations. We here provide reward for objectives in \S~\ref{subsubsec:intervention}:

(1) Promote cross-viewpoint interactions: Inspired by the commonly used evaluation metrics~\cite{törnberg2023simulatingsocialmediausing, mou-etal-2024-unveiling, yang2024oasisopenagentsocial}, the reward $r_i^t$ for arm $x_i^t$ balances: (1) engagement across opposing stances, (2) penalizing toxic interactions, and (3) preserving overall user engagement. Formally, given a post-user arm $x_i^t$ and the reaction $o_i^{t+1}$ from the receiver agent (see \S \ref{subsubsec:behavior}),  the reward for arm $x_i^t$ is
\begin{equation}
r_i^t = \frac{|s^t(u_s)-s^t(u_r)|}{2} h(o_i^{t+1}) \max\big(0, 1 - \mu,\tau(o_i^{t+1})\big),
\end{equation}
balancing stance divergence between sender $u_s$ and receiver $u_r$, penalizing toxic reactions $\tau(o_i^{t+1}) \in [0,1]$ (via Perspective API~\cite{lees2022new}), and weighting the reaction type by engagement through $h(o_i^{t+1})$.

(2) Mitigate misinformation: Similarly, we compute the reward based on the environment’s reaction. Specifically, for arm $x_i^t$,
\begin{equation}
r_i^t = \text{mis}^{t-1}(u) - \text{mis}^{t}(u),
\end{equation}
where $\text{mis}^t(u)$ indicates whether user $u$ is misinformed at round $t$.

\vpara{Optimization Process:}  In the bandit optimization process, we draw inspiration from the works~\cite{kassraie2022graph,ban2021ee} and consider both the exploitation and exploration framework, as detailed below.

For the exploitation aspect, actions are made based on the knowledge or experience already acquired. We here use a neural network $g$ to learn the mapping $g_{\theta^t}: x_i^t \to r_i^t $, where the context of the arms is mapped to the reward with learnable parameters $\theta^t$. After executing an arm $x_i^t$, we receive the reward $r_i^t$ at the next simulation epoch $t+1$. Therefore, we perform gradient descent to update $\theta^t$ based on the collected feedback $\left\{x_i^t, r_i^t\right\}$.

In addition to exploiting the contexts, model should explore new possibilities in unknown environments to discover potentially better policies. Inspired by~\cite{ban2021ee}, we utilize neural network $\hat{g}_{\phi^t}$ to estimate the potential gain in terms of reward for exploration, where potential gain measures the discrepancy between the observed reward and the predicted reward $r_i^t-g_{\theta^t}(x_i^t)$.  A large positive potential gain indicates that the arm is more under-explored, while a small potential gain suggests an overestimation of the reward, making it less suitable for exploration.

Since the potential gain is bounded by the gradient of the predicted reward $\nabla_{\theta_t} g(\mathbf{x}_i^t)$ in~\cite{ban2021ee}, the gradient of the predicted reward is used as input to measure the potential gain $\hat{g}_{\phi^t}: \nabla_{\theta_t} g(x_i^t)\rightarrow r_i^t-g_{\theta^t}(x_i^t)$. After executing an arm $x_i^t$, we receive the reward $r_i^t$ at the next simulation epoch $t+1$. Therefore, we perform gradient descent to update $\phi^t$ based on the collected feedback $\left\{\nabla_{\theta_t} g(x_i^t), r_i^t-g_{\theta^t}(x_i^t)\right\}$. Finally, to balance exploitation and exploration, we output score $g_{\theta^t}(x_i^t) + \hat{g}_{\phi^t}(\nabla_{\theta_t} g(x_i^t))$, which is used to select the arms by ranking.
% \vspace{-0.3in}
\section{Experiments}\label{sec:exp}
In this section, we evaluate \ours~in two stages: first, the realism of the social simulation, and second, the effectiveness of adaptive intervention policy (Code is available at \url{https://github.com/renH2/PolicySim}.). Our experiments address two key questions:

\vpara{1. How valid is the simulation generated by \ours?} We propose several metrics to identify suitable agents for social simulation and validate \ours~at both micro- and macro-levels.

\vpara{2. Can the adaptive intervention policy effectively optimize platform outcomes?} We test our adaptive intervention policy under various objectives, showing \ours~can leverage environmental feedback to optimize policies.

% \vspace{-0.1in}
\subsection{Experimental Setup}\label{subsec:setup}
\subsubsection{Datasets.} In our experiments, we use real-world social media datasets: TwiBot-20 dataset~\cite{feng2021twibot}. TwiBot-20, collected from July to September 2020, comprises 229K users, 33.5M tweets, and 456K follow links. We extract social activity events and relevant user groups while preserving their relational structures. In addition, we further conduct experiments on the Weibo dataset~\cite{ma2016detecting}. Detailed statistics of the dataset are provided in Appendix~\ref{app:exp-setup}, and the experiments conducted on the Weibo dataset are presented in Appendix~\ref{app:exp}.

\begin{table*}[!th]
\centering
\resizebox{2.1\columnwidth}{!}{
\renewcommand{\arraystretch}{1.7}
\begin{tabular}{c|cc|c|c|ccc}
\toprule
\multirow{2}{*}{Method} & \multicolumn{2}{c|}{Content quality} & Behavior alignment & Self-consistency & \multicolumn{3}{c}{Social capability}  \\ \cmidrule{2-8} 
&\textit{BERTScore F1} $\uparrow$  & \textit{BertSim}$\uparrow$  & \textit{Accuracy} $\uparrow$ & \textit{Accuracy} $\uparrow$ & \textit{Engagement} $\uparrow$ & \textit{Robustness} $\uparrow$ &  \textit{Suitability} $\uparrow$ \\ \midrule
Random & \makecell{28.55 \textcolor{gray}{\small ±4.18}}&\makecell{74.42 \textcolor{gray}{\small ±12.08}} &\makecell{36.11 \textcolor{gray}{\small ±25.30}} &\makecell{21.20 \textcolor{gray}{\small ±1.65}}& \makecell{2.65 \textcolor{gray}{\small ±0.55}}& \makecell{2.31 \textcolor{gray}{\small ±0.54}}& \makecell{2.11 \textcolor{gray}{\small ±0.14}} \\ \cline{1-8}
% Qwen2.5-0.5B-Instruct & 46.64±23.99&81.38±9.03 & 47.78±22.74&24.3±19.52& 2.95±0.45& 2.43±0.57&47.22±0.50\\ 
% GLM4-light & 46.01±10.35&81.95±13.64&0.4833±0.2764&0.472±0.2764&3.30±0.51&2.70±0.57&63.19±0.48\\
GLM4-light~\cite{glm2024chatglm} & \makecell{46.01 \textcolor{gray}{\small ±10.35}} &  \makecell{81.95 \textcolor{gray}{\small ±13.64}} & \makecell{48.33 \textcolor{gray}{\small ±27.64}} & \makecell{47.20 \textcolor{gray}{\small ±27.64}} & \makecell{2.91 \textcolor{gray}{\small ±0.40}} & \makecell{2.70 \textcolor{gray}{\small ±0.57}} & \makecell{57.64 \textcolor{gray}{\small ±0.50}} \\
% Llama-3-8B-Instruct& 46.32±13.07&85.22±6.81& 0.5236±0.2621&0.456±0.2595&3.16±0.49&	3.15±0.81	&	71.85±0.45\\
Llama-3-8B-Instruct~\cite{llama3modelcard} & \makecell{46.32 \textcolor{gray}{\small ±13.07}} & \makecell{85.22 \textcolor{gray}{\small ±6.81}} & \makecell{52.36 \textcolor{gray}{\small ±26.21}} & \makecell{45.60 \textcolor{gray}{\small ±25.95}} & \makecell{3.16 \textcolor{gray}{\small ±0.49}} & \makecell{3.51 \textcolor{gray}{\small ±0.50}} & \makecell{\underline{71.85} \textcolor{gray}{\small ±0.45}} \\
Qwen2.5-0.5B-Instruct~\cite{qwen2.5} & \makecell{46.64 \textcolor{gray}{\small ±23.99}} &  \makecell{81.38 \textcolor{gray}{\small ±9.03}} &  \makecell{47.78 \textcolor{gray}{\small ±22.74}} &  \makecell{24.30 \textcolor{gray}{\small ±19.52}} &  \makecell{2.95 \textcolor{gray}{\small ±0.45}} &  \makecell{2.43 \textcolor{gray}{\small ±0.57}} & 
\makecell{47.22 \textcolor{gray}{\small ±0.50}} \\ 
% Qwen2.5-3B-Instruct & 48.26±12.57&85.91±6.35 & 60.56±26.87&40.4±26.98&3.17±0.47& 2.65±0.61&52.41±0.50\\
Qwen2.5-3B-Instruct & \makecell{48.26 \textcolor{gray}{\small ±12.57}} & \makecell{85.91 \textcolor{gray}{\small ±6.35}} & \makecell{60.56 \textcolor{gray}{\small ±26.87}} & \makecell{40.40 \textcolor{gray}{\small ±26.98}} & \makecell{3.17 \textcolor{gray}{\small ±0.47}} & \makecell{2.65 \textcolor{gray}{\small ±0.61}} & \makecell{52.41 \textcolor{gray}{\small ±0.50}} \\
% Qwen2.5-7B-Instruct& 49.48±11.77&85.52±6.63&0.5056±0.256&0.512±0.2998&3.29±0.47&2.71±0.61&63.83±0.48\\
Qwen2.5-7B-Instruct & \makecell{49.48 \textcolor{gray}{\small ±11.77}} & \makecell{85.52 \textcolor{gray}{\small ±6.63}} & \makecell{50.56 \textcolor{gray}{\small ± 25.60}} & \makecell{51.20 \textcolor{gray}{\small ±29.98}} & 
\makecell{3.29 \textcolor{gray}{\small ±0.47}} & \makecell{2.71 \textcolor{gray}{\small ±0.61}} & 
\makecell{63.83 \textcolor{gray}{\small ±0.48}} \\ \cline{1-8}
\ours-$\phi$ & \makecell{45.16 \textcolor{gray}{\small ±14.91}} & \makecell{80.15 \textcolor{gray}{\small ±8.23}} & \makecell{58.33 \textcolor{gray}{\small ±29.95}} & \makecell{27.20 \textcolor{gray}{\small ±21.82}} & \makecell{3.04 \textcolor{gray}{\small ±0.42}} & \makecell{2.52 \textcolor{gray}{\small ±0.57}} & \makecell{55.17 \textcolor{gray}{\small ±0.50}} \\
\ours-SFT & \makecell{52.66 \textcolor{gray}{\small ±15.53}} & \makecell{86.77 \textcolor{gray}{\small ±7.55}} & \makecell{54.44 \textcolor{gray}{\small ±22.91}} & \makecell{\underline{56.40} \textcolor{gray}{\small ±25.83}} & \makecell{3.00 \textcolor{gray}{\small ±0.42}} & \makecell{2.42 \textcolor{gray}{\small ±0.53}} & \makecell{44.83 \textcolor{gray}{\small ±0.50}} \\
\ours-DPO & \makecell{47.95 \textcolor{gray}{\small ±13.24}} & \makecell{83.20 \textcolor{gray}{\small ±8.13}} & \makecell{53.89 \textcolor{gray}{\small ±24.41}} & \makecell{50.40 \textcolor{gray}{\small ±25.37}} & \makecell{3.14 \textcolor{gray}{\small ±0.47}} & \makecell{2.67 \textcolor{gray}{\small ±0.58}} & \makecell{61.27 \textcolor{gray}{\small ±0.49}} \\ \cdashline{1-8}[1pt/1pt]
\ours & \makecell{\underline{58.05} \textcolor{gray}{\small ±15.96}} & \makecell{\underline{88.06} \textcolor{gray}{\small ±6.32}} & \makecell{\underline{65.56} \textcolor{gray}{\small ±19.71}} & \makecell{56.00 \textcolor{gray}{\small ±25.61}} & \makecell{\underline{3.20} \textcolor{gray}{\small ±0.44}} & \makecell{\underline{2.73} \textcolor{gray}{\small ±0.61}} & \makecell{59.44 \textcolor{gray}{\small ±0.49}} \\ \bottomrule
\end{tabular}
}
\caption{Micro-level performance of social simulation across different backbones and \ours~on the TwiBot-20 dataset. Underline indicates the best result and results are averaged over five runs with standard deviations.}
\label{tab:micro_results}
\vspace{-0.2in}
\end{table*}

\begin{table*}[!ht]
\centering
\resizebox{1.3\columnwidth}{!}{
\renewcommand{\arraystretch}{1.3}
\begin{tabular}{c|ccc|c}
\toprule
\multirow{2}{*}{Method} & \multicolumn{3}{c|}{Objective 1} & Objective 2  \\ 
& Stance & Toxicity $\downarrow$  & Cross interactions  $\uparrow$ & Misinformation ratio $\downarrow$  \\ \midrule
Origin & 0.014 (0.47)& 0.0556& 0.04 & 40\%\\ \cdashline{1-5}[1pt/1pt]
$\epsilon$-greedy &0.184 (0.42) & 0.0426&0.14 & 26\%   \\
UCB &  0.026 (0.34)&0.0628 &0.50 & 30\% \\ \cdashline{1-5}[1pt/1pt]
% \ours-TS &\\ 
\ours &0.376 (0.48)&\underline{0.0386} &\underline{0.56}&\underline{24\%}
\\
\bottomrule
\end{tabular}
}
\caption{Performance of different intervention policies across different objectives. Underline indicates the best performance.}
\label{tab:adaptive-result}
\vspace{-0.1in}
\end{table*}

\subsubsection{Simulation evaluation metric.}
The evaluation metrics for the simulation encompass both \textbf{micro-level} and \textbf{macro-level} perspectives. At the \textbf{micro-level}, we focus on assessing whether social agents’ decision-making behaviors align with real-world patterns. Specifically, we evaluate four aspects:

\begin{itemize}[leftmargin=12pt]
\item \textbf{Content quality.}  We assess whether agents can produce realistic content by comparing generated posts with real user posts using multiple textual similarity metrics, including \textit{BERTScore F1} and \textit{BertSim} (BERT-based cosine similarity), to capture both semantic and lexical closeness.  
\item \textbf{Behaviour alignment.} We measure the ability of agents to replicate human-like behavioral actions (e.g., posting, retweeting) through prediction accuracy.
\item \textbf{Self-consistency.} We measure whether agents can correctly identify their own generated posts by prediction accuracy, reflecting the self-consistency of  behaviors.
\item \textbf{Social capability.} We further employ large language models as evaluators (\textit{LLM-as-a-Judge}) to rate each agent on three dimensions: (1) \textit{Engagement}: the agent's ability to participate in natural, meaningful, and contextually appropriate interactions, such as replying, expressing opinions, and conveying emotions; (2) \textit{Robustness}: the agent's capacity to maintain relevance, coherence, and naturalness across diverse social contexts, including discussions, debates, humor, and trending topics; (3) \textit{Suitability}: an overall measure of the agent’s behavioral realism, reflecting how convincingly the agent mimics authentic human activity.
\end{itemize}
Notably, \textit{Engagement} and \textit{Robustness} are measured from 1 to 4 ordinal scale, whereas \textit{Suitability} is evaluated from 0 to 100 on continuous scale. 

At the \textbf{macro level}, we measure how distribution of stances evolves over time to assess whether agents reproduce realistic opinion dynamics. In addition, we examine how intervention policies affects network-level phenomena such as polarization. 

\subsubsection{Backbones and Baselines.} 
We evaluate \ours~through both simulation and adaptive intervention experiments. For simulation, \ours~is built upon five open-source LLMs used in a prompt manner, including GLM4-light, Llama-3-8B-Instruct, and the \textsc{Qwen2.5} series. To examine the effect of each training component, we conduct ablations: (1) \ours-$\phi$ (without user profile generation); (2) \ours-SFT (only supervised fine-tuning); and (3) \ours-DPO (only DPO without SFT initialization). For intervention, we compare with$\epsilon$-greedy and UCB bandit baselines.

\subsubsection{Implementation details.} In practice, we set the attenuation coefficient $\alpha$ in Eq.(\ref{eq:stance_update}) to 0.8, and use $\lambda = 1$, $k = 1$, $\mu = 4$, and $\beta = 0.5$ in all experiments. For the macro-level results in \S~\ref{subsec:simulation-result} and Objective 1 in \S~\ref{subsec:adaptive-result}, we use the social topic \emph{Anti-abortion Legislation} as the simulation context, collecting 10 trigger news items from late June 2022 (after the overturning of \emph{Roe v.Wade}) via \emph{newsapi.ai} and summarizing them with GPT-4o. For Objective 2, we initialize 20\% of users to post misinformation—``\#wakeupamerica who needs a \#gun registry when \#obama has all your personal information''—and observe its spread in the simulated environment.

Owing to its strong capability on social interaction, The LLM we employ to power the user agents is Qwen2.5-3B-Instruct~\cite{qwen2.5}, with a maximum context length of 32,768 tokens. For model training, we use a LoRA adaptation of rank 64 to finetune the base model with a learning rate of $1\times 10^{-6}$, batch size of 256. In the DPO stage, we set the temperature coefficient  $\beta = 0.1$, learning rate $5\times 10^{-7}$ and sample size $J = 3$. Both stages are trained for up to 10 epochs until convergence. All experiments were conducted on 12 NVIDIA A100-PCIE-40GB GPUs. More details can be found in Appendix~\ref{app:exp-setup}. 

% \vspace{-0.2in}
\subsection{Simulation Results}\label{subsec:simulation-result}
\subsubsection{Micro-level evaluation.} As shown in Table~\ref{tab:micro_results}, we first report the micro-level simulation results of \ours. In terms of \textbf{content quality}, LLM-based agents effectively reproduce the linguistic and semantic characteristics of real Twitter posts, achieving a high \textit{BertSim} score and \textit{BERTScore F1}, demonstrating its ability to generate realistic, human-like text. For \textbf{behaviour alignment}, \ours~improves the accuracy of reproducing human behavioral patterns by 8.26\% over random and backbone baselines, highlighting the effectiveness of modeling user interaction during agent training. \ours~also exhibit strong self-consistency, with an accuracy gain of 10.15\%, indicating coherence between generated content and underlying behavioral preferences. Regarding \textbf{social capability}, \ours~achieves superior scores in \textit{Engagement} and \textit{Robustness}, demonstrating its ability to emulate context-aware and socially coherent user behaviors. Notably, Llama-3-8B-Instruct performs better on \textit{Suitability}, likely due to its instruction tuning or pre-traning data that promote more socially appropriate responses.

As for the comparison between \ours~and other baseline models, the results reveal that the \textsc{Qwen} series exhibits consistent improvements with increasing model scale across most metrics (except for \textit{BertSim} and accuracy of behavior alignment), confirming the applicability of scaling laws in social simulation tasks. Besides, the performance of \ours$-{\phi}$, which removes the user profile module, declines apparently compared to \ours, underscoring the pivotal role of user profiling in shaping personalized agent behaviors. Finally, the inferior performance of \ours-DPO relative to \ours-SFT and \ours~indicates that supervised fine-tuning serves as a necessary foundation for acquiring the core structure, style, and knowledge of social interactions before applying DPO to achieve optimal performance.

\subsubsection{Macro-level evaluation.} We simulate real-world dynamics by chronologically injecting trigger news events into the environment. Each news event is designed to shift public opinion. For example, in rounds 0 and 4, agents are exposed to news headlines “The Supreme Court overturns Roe v. Wade” and “The Biden administration announced plans to continue covering abortion medication,” respectively. In round 7, large-scale protests demanding federal action to restore abortion rights are introduced. Agents supporting Anti-Abortion Legislation are labeled as having a positive stance, while opponents are labeled as negative.

\begin{figure}[h]
    \centering
    \includegraphics[width=0.47\columnwidth]{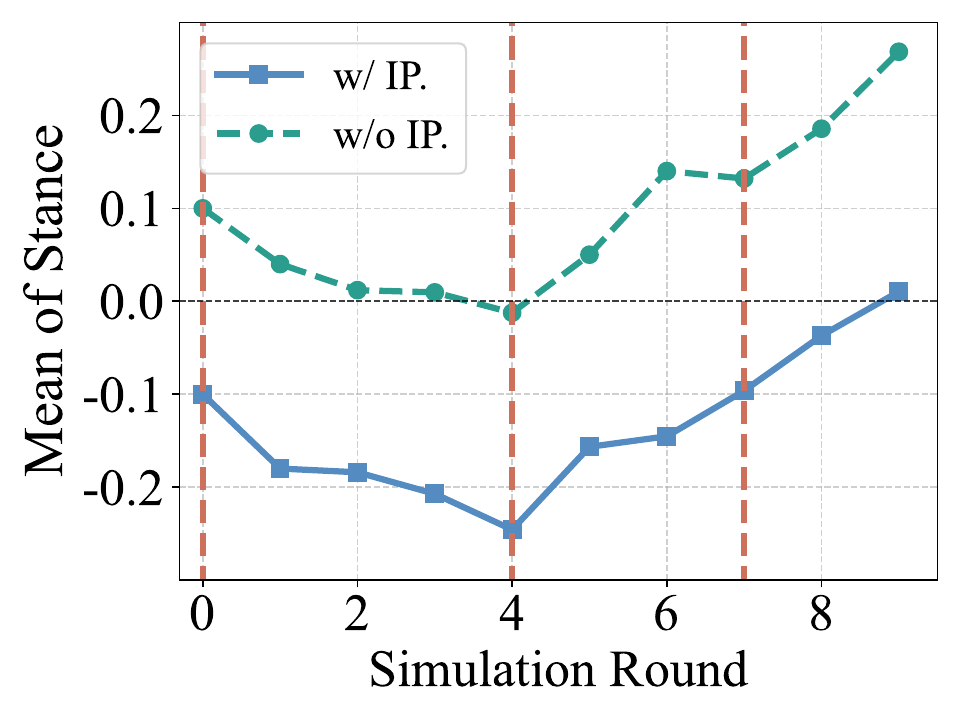}
    \hfill
    \includegraphics[width=0.47\columnwidth]{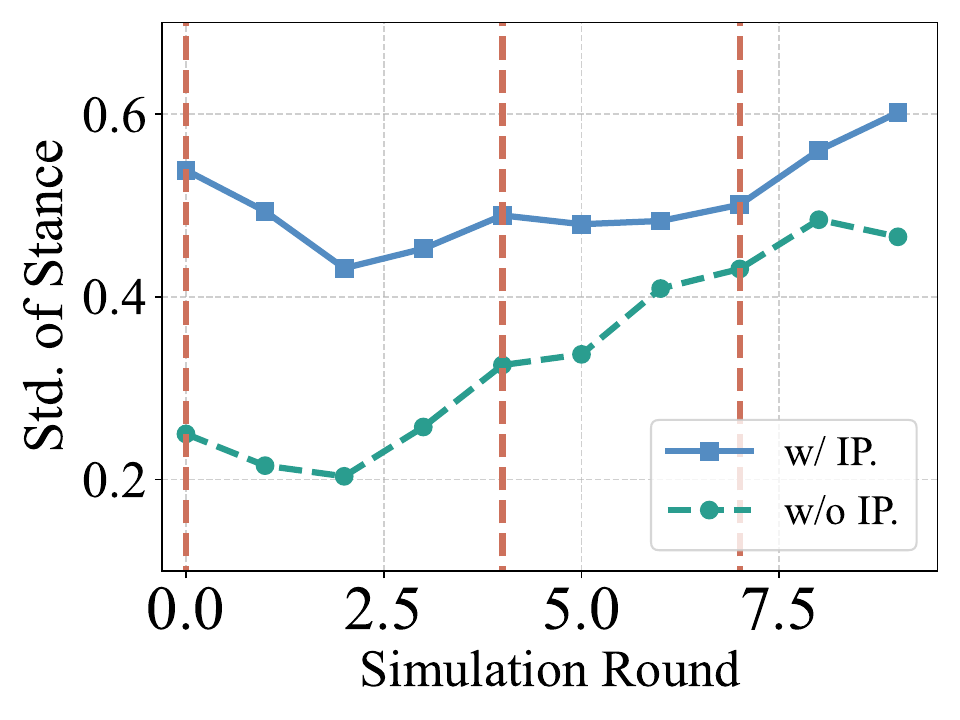}
    \vspace{-0.1in}
    \caption{Mean and Std. of stance score by different intervention policies under trigger news. w/ IP. and w/o IP. indicate whether an intervention policy is applied.}
    \label{fig:macro}
    \vspace{-0.2in}
\end{figure}

As shown in Figure~\ref{fig:macro}, the mean stance across agents exhibits a rapid initial decline, followed by a gradual increase, mirroring the expected public opinion trajectory. This validates the realism of our simulation in capturing macro-level stance dynamics. Moreover, the std. of stance scores increases over rounds, indicating a polarization effect as agents form more extreme opinions. Notably, when the intervention policy (i.e., recommender system) is applied, polarization intensifies: users are increasingly exposed to homogeneous content, reinforcing existing beliefs and amplifying divisions.

\vspace{-0.1in}

\subsubsection{Scalability.} 
\begin{figure}[h]     
    \centering
    \includegraphics[width=0.5\columnwidth]{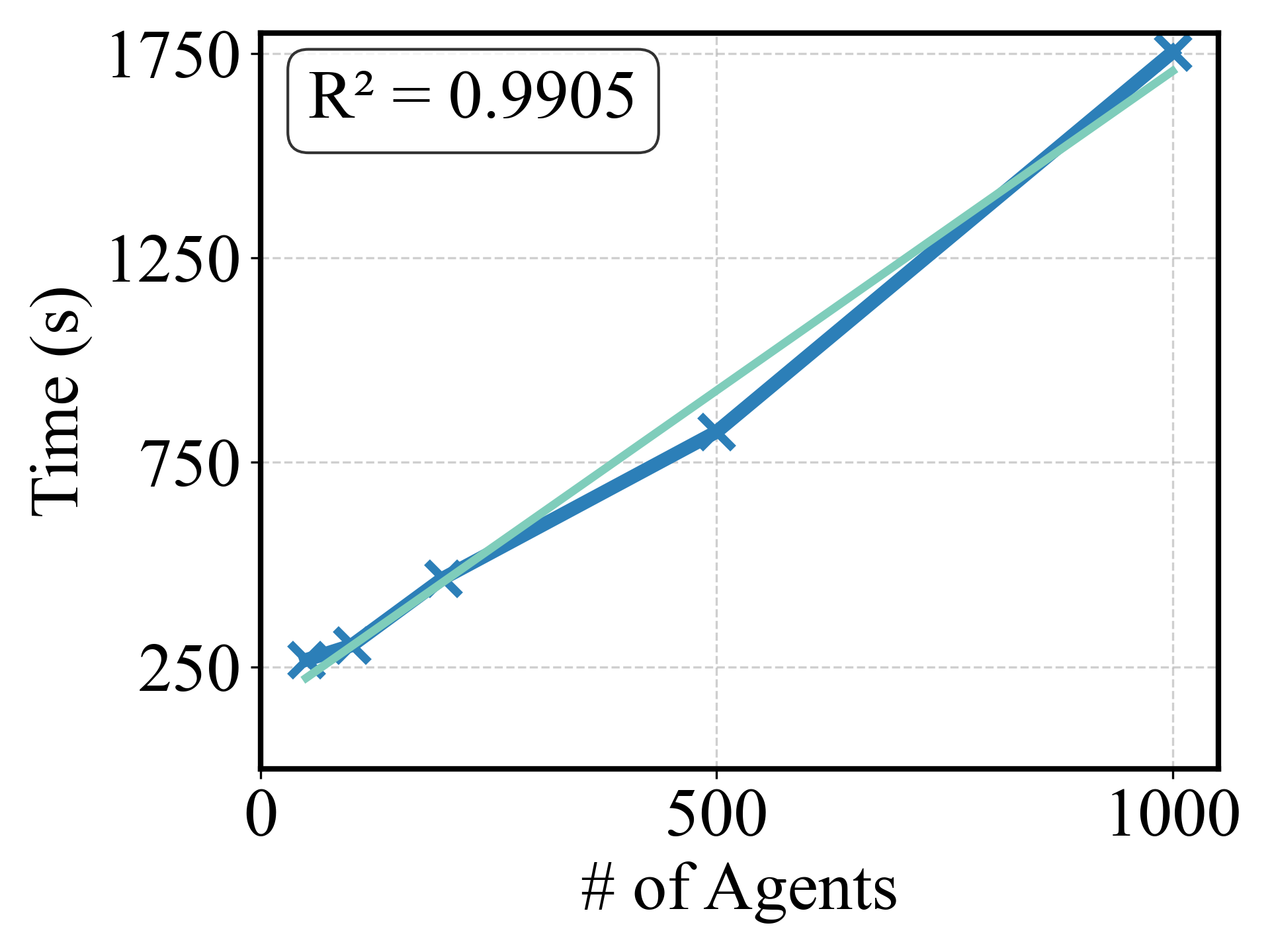}
    \vspace{-0.1in}
    \caption{Scalability of \ours~on the TwiBot-20 dataset, showing linear runtime growth with agent scale ($r = 0.9904$).}\label{fig:scalability}
\end{figure}

Figure~\ref{fig:scalability} illustrates the execution time across different numbers of agents over 10 simulation rounds. The overall runtime is primarily determined by the latency of LLM inference (or API calls) and the execution of the intervention policy. As shown, the total computational cost increases approximately linearly with the number of agents, demonstrating the scalability and efficiency of our system under larger-scale simulations.

\vspace{-0.2in}

\subsection{Adaptive Intervention Policy Results}\label{subsec:adaptive-result}

\vpara{Intervention result} 
We evaluate the effectiveness of \ours~under different \textit{intervention objectives}, as shown in Section~\ref{subsubsec:intervention}. 

\vpara{Objective 1: Promoting Cross-Viewpoint Interaction.} This objective aims to promote \textit{cross-viewpoint interactions} without increasing \textit{toxicity} by adjusting the recommender system. In this setting, we measure (i) the average and standard deviation of stance scores, (ii) the overall toxicity level on the platform, and (iii) the ratio of cross-stance interactions among all interactions. 

\vpara{Objective 2: Mitigating Misinformation Propagation.}
This objective focuses on mitigating the impact of \textit{misinformation} by controlling the exposure mechanism. Specifically, we initialize 20\% of users to post misinformation and then measure the propagation ratio within the simulated environment. 

As shown in Table~\ref{tab:adaptive-result}, for \textbf{Objective 1}, \ours~significantly increases the proportion of cross-stance interactions while simultaneously reducing overall toxicity. In contrast, $\epsilon$-greedy and UCB also enhance cross-stance engagement but do so less precisely, often triggering conflicts between opposing stances and thereby increasing toxicity. The reduced standard deviation of stance scores further indicates that \ours~effectively enhances viewpoint diversity. For \textbf{Objective 2}, by controlling the exposure mechanism, \ours~successfully limits misinformation diffusion according to environment feedback, demonstrating its adaptability and broad applicability across different intervention objectives.
% \vspace{-0.1in}
\section{Related Work}
\vpara{Multi Agent Social Simulation.} LLM-driven multi-agent systems have emerged as a powerful paradigm for social simulation, overcoming the limitations of early rule-based or psychologically inspired models~\cite{schwartz2013personality, huang2019hierarchical, deffuant2000mixing,luo2024llm} that struggled to capture the complexity and adaptability of human behavior. A pioneering effort, \emph{social simulacra}~\cite{park2022social}, introduced autonomous agents capable of human-like reasoning, memory, and decision-making in large-scale social computing systems. Subsequent frameworks~\cite{chen2023agentverse, liu2024lmagent, mou2024unveiling, yang2024oasisopenagentsocial} extended this line of work by integrating multi-modal information, supporting diverse application scenarios, and enabling scalable deployment of multiple LLM-based agents within social networks. These systems have been applied to collaborative planning and discussion, testing social science theories~\cite{chuang2023wisdom}, simulating realistic communities~\cite{park2023generative,yang2024oasisopenagentsocial}, modeling opinion dynamics~\cite{chuang2023opinion}, and even macroeconomic patterns~\cite{li2023econagent}. Despite these advances, existing work remains limited in simulation authenticity and has rarely leveraged simulations to optimize models for real-world applications, motivating approaches that enhance both the fidelity and practical utility of social simulations.

% Social Intervention addresses issues arising from social media platforms. While the algorithms adopted by the platforms enhance user engagement by curating personalized content, they have also been criticized for fostering phenomena such as \emph{echo chambers} and \emph{polarization}

\vpara{Social Intervention.} Social Intervention refers to the concept that arises alongside the development of social media platforms. While the algorithms adopted by the platforms enhance user engagement by curating personalized content, they have also been criticized for fostering phenomena such as \emph{echo chambers} and \emph{polarization}~\cite{cinelli2021echo,bail2018exposure}. Social intervention aims to balance engagement with the promotion of healthier, more constructive interactions. For instance, some platforms have experimented with \emph{nudging} mechanisms~\cite{jesse2021digital}, where users are prompted to engage with content critically before sharing it. Others have leveraged graph-based diversification techniques~\cite{xu2025user, yuan2025tree, huang2024can, xu2023better, huang2024measuring} to introduce heterogeneity~\cite{yuan2024unveiling,sun2022beyond} in recommended content, preventing the formation of insular communities~\cite{zheng2021dgcn,he2020lightgcn}. Several works have utilized the llm-based environment to study the impact of different intervention policies~\cite{törnberg2023simulatingsocialmediausing, mou2024unveiling}. However, most existing studies focus on testing fixed intervention based on predefined objectives. In contrast, our work treats intervention mechanisms as learnable modules. By leveraging feedback from the environment, we optimize intervention policies via reinforcement learning to achieve desired objectives, providing insights for real-world strategies.
% \vspace{-0.1in}
\section{Conclusion} 
This work presents PolicySim, an LLM-driven social simulation sandbox that enables proactive assessment and optimization of social platform interventions before deployment. On the simulation side, we introduce intervention modules to construct realistic social media environments and, for the first time, adapt LLMs to social agents through training process instead of prompt engineering. To achieve the predefined objectives, PolicySim balances exploration and exploitation in adaptive policy learning via a contextual bandit algorithm enhanced with a message passing mechanism. Extensive experiments demonstrate the framework’s effectiveness in simulation and the generality of the scenarios, underscoring its potential as a novel paradigm for intervention policy design.

\section*{Acknowledgements}
This work is supported by NSFC (No.\ 62206056, No.\ 62322606, No.\ 62441605), the CIPSC-SMP-Zhipu Large Model Cross-Disciplinary Fund, and a collaboration funding by MYbank, Ant Group.

\newpage

\bibliography{reference}
\bibliographystyle{ACM-Reference-Format}

% \clearpage
%\onecolumn
\appendix
\newpage
\section{Appendix}
\subsection{Framework} \label{app:framework}

In this section, we detail the pseudocode for the algorithm behind \ours. We outline the overall procedure of \ours~as follows Algorithm~\ref{alg:our}.

\begin{algorithm}[h]
    \caption{Framework of \ours~sandbox}
    \begin{algorithmic}[1]
        \REQUIRE Twitter, historical post, total simulation round $T$, total user number $N$.
        \ENSURE Initialization for $X^{\prime}$ and  $Y^{\prime}$.
        \STATE Initialize agents profile.
        \STATE Initialize agents score $s^0$.
        \STATE Initialize $G$ of the agents' relationship.
        \FOR{$t$ in $0, 1,\cdots ,T$}
            \FOR{$i$ in $0, 1,\cdots , N$}
            \STATE Agent $u_i$ generates response based on its profile $p(u_i)$, context and memory.
            \ENDFOR
            \STATE Obtain the $t-1$ round reaction.
            \STATE Obtain the reward $r^{t-1}$.
            \STATE Conduct recommendation by ranking the predicted reward $g_{\theta^t}(x_i^t) + \hat{g}_{\phi^t}(\nabla_{\theta_t} g(x_i^t))$.
            \STATE Update $g_{\theta^t}$ by $\left\{x_i^{t-1}, r_i^{t-1}\right\}$.
            \STATE Update $\hat{g}_{\phi^t}$ by $\left\{\nabla_{\theta_t} g(x_i^{t-1}), r_i^t-g_{\theta^{t-1}}(x_i^t)\right\}$.
            \STATE Update short-term memory and long-term memory.
            \STATE Update stance score $s^t$.
        \ENDFOR
    \end{algorithmic}\label{alg:our}
\end{algorithm}

\vspace{-0.1in}
\subsection{Addition Experimental Setup}\label{app:exp-setup}

\vpara{Detailed statistics of Twitter dataset.}
For the data preprocess, we first filtered relevant non-robot users from the Twibot-20 dataset\cite{feng2021twibot}, focusing on those associated with the "Politics" domain. The filtering process involved selecting users who have both tweets and neighbors. We collected key profile features such as user name, screen name, description, account creation date, location, follower count, friends count, and favorite count. Additionally, we limited the tweet content to a maximum of 20 records. For the social network relationships, we extracted follower and following information to build a directed graph representing user connections. Below is the statistics of the datasets used in our experiments.

\begin{table*}[!ht]
% \fontsize{9pt}{10pt}\selectfont
\setlength \tabcolsep{2pt}
\renewcommand{\arraystretch}{1}
\centering
{
\begin{tabular}{lcccccccc}
 \toprule
  &\# Node &\# Edges &\# of tweets&Max Degree&Average Degree&Density & Average Cluster Modularity &Average Length of tweet \\ \hline
Value &924 &302 &20,061 & 14.0 &0.65 &3.54e-4
 &0.96 &115.61 \\   \bottomrule
\end{tabular}
}
\caption{Dataset statistics for Twibot-20 sampled dataset.}\label{tab:statistic}\vspace{-0.2in}
\end{table*}
\vpara{Specific format of Twitter metadata}
We process the metadata into the following format, with the information anonymized as listed below.
\begin{tcolorbox}[colback =red!5!white, colframe =red!75!black, breakable, title = MetaData]
\begin{verbatim}
"ID": "34209XXXX",
"profile": {
    "name": "XXX",
    "screen_name": "XXX_XXX",
    "description": "XXX ",
    "created_at": "Thu Aug 13 21:38:42 2015 ",
    "followers_count": "8856 ",
    "friends_count": "1182 ",
},
"tweet": ["On the birthday of our country,..."],
"neighbor": {
    "following": ["74605XXXX",...],
    "follower": ["2006XXXX",...]
}
\end{verbatim}
\end{tcolorbox}

\vpara{Design for extracting four high-level information}
\begin{tcolorbox}[colback =red!5!white, colframe =red!75!black, breakable, title = Information Extraction]
\begin{verbatim}
Assume you are playing the role of a user 
in a social network.
{Agengt ID}, {User Info}
Historical Tweets:[Tweet1:...., Tweet2:...],
Generate a concise user persona covering: 
1. The likely identity of the user: .....
2. The user’s main areas of focus 
3. The user’s posting style
4. The user’s interaction behavior 
### Output format (JSON): 
{
    "synthetic_profile":
}
\end{verbatim}
\end{tcolorbox}

\vpara{Micro evaluation metric via LLM as judge prompt}
\begin{tcolorbox}[colback =red!5!white, colframe =red!75!black, breakable, title = Micro Evaluation]
\begin{verbatim}
You are an evaluator.  
We want to determine whether a given agent is 
suitable to act as a social agent.  

You are provided with:  
- **User profile information** 
- **Historical posts from this user**
- **Agent's generated responses **

Please evaluate the agent from :  
1. **Social Engagement Ability**: ...
2. **Identity Consistency**:...
3. **Robustness** 
### Output format (JSON):  
{
"Social Engagement Ability": "score (1–5) ",
"Identity Consistency": "score (1–5) ",
"Robustness": "score (1–5)",
"Final Judgment": "Yes/No "
}
\end{verbatim}
\end{tcolorbox}

\vpara{Prompt for memory designs}
Below are some of our prompts for short-term memory and long-term memory.

\begin{tcolorbox}[colback =red!5!white, colframe =red!75!black, breakable, title = Short Term Memory]
\begin{verbatim}
Assume you are a user in a social network
{synthetic_profile}
Message: {message.to_string()}
Memory: {memory[-1].to_string()}
Identify the key parts in the 
message that are most relevant or important.
### Output format (JSON): 
{
    "short_term_memory":
}
\end{verbatim}
\end{tcolorbox}

\begin{tcolorbox}[colback =red!5!white, colframe =red!75!black, breakable, title = Long Term Memory]
\begin{verbatim}
Assume you are a user in a social network
{synthetic_profile} 
Message: {message.to_string()}
Short Memory: {short_term_memory}
Summarize your long-term memory about this message
### Output format (JSON): 
{
    "long_term_memory":
}
\end{verbatim}
\end{tcolorbox}

\vpara{Prompt for calling different actions}

\begin{tcolorbox}[colback =red!5!white, colframe =red!75!black, breakable, title = Actions calling]
\begin{verbatim}
The TOPIC for this simulation is "{topic}"
At the very moment, you have got several latest news 
Trigger news: {News-...}
User memory: {memory_review}
Message: {recommend_message}
Have you followed the message sender: {True/False}
Generate a reaction to this message by calling:
- **do_nothing()** 
- **post(content)**
- **retweet(content)**
......
### Reasoning Guidence
...
### Response format(JSON):
{
    [
    {"action": "retweet", "content": "......"},
    {"action": "follow"}...
    ]
}
\end{verbatim}
\end{tcolorbox}

\vpara{Additional implementation details.} 
We further elaborate on the hyper-parameters used and the running environment. Both exploration and exploitation models consist of a two-layer fully connected network with embedding size 768, hidden layer size 64, trained with the Adam optimizer. All other parameters are the default settings from the HuggingFace library~\cite{wolf2020huggingfacestransformersstateoftheartnatural}. As for the running environment, our model is implemented under the following software setting: openai version 1.52.0, Pytorch version 2.0.2+cu121, CUDA version 12.4, networkx version 3.2.1, transformers version 4.46.0, numpy version 1.26.4, Python version 3.10.15.

\subsection{Additional Experimental Results}\label{app:exp}
\vpara{Robustness of \ours~under different LLM hyperparameters.}
We further evaluate the robustness and generalizability of our framework by analyzing generation diversity under different temperature settings ($\tau \in [0.4, 1.0]$) using the Hunyuan-Lite backbone. The results are summarized in Table~\ref{tab:robustness}.
\vspace{-0.1in}
\begin{table}[!h]
    \centering
    \setlength{\tabcolsep}{1.3pt}
    \renewcommand{\arraystretch}{1.05}
    \resizebox{0.9\columnwidth}{!}{
    \begin{tabular}{c|ccccccc}
    \toprule[1pt]
    Temperature ($\tau$) & 0.4 & 0.5 & 0.6 & 0.7 & 0.8 & 0.9 & 1.0 \\ \hline
    PolicySim & 0.7584 & 0.7649 & 0.7614 & 0.7632 & 0.7675 & \textbf{0.7769} & 0.7523 \\ 
    \bottomrule[1pt]
    \end{tabular}}
    \caption{Robustness evaluation of \ours~under varying temperature parameters ($\tau$) on Hunyuan-Lite. Moderate temperatures ($\tau = 0.9$) achieve the best balance between generation diversity and semantic coherence.}
    \label{tab:robustness}
    \vspace{-0.3in}
\end{table}

We observe that moderate temperatures (e.g., $\tau = 0.9$) optimally balance diversity and coherence, better approximating human creative tendencies. In contrast, lower temperatures yield overly deterministic and repetitive responses, while higher temperatures induce semantic drift and reduce overall quality, as the model tends to produce incoherent outputs.

\vpara{Simulation result on Weibo dataset.} When transferring the LLM trained on the TwiBot dataset to the Weibo dataset, our approach maintains strong performance and surpasses the directly used Qwen2.5-3B-Instruct baseline, highlighting the cross-platform generalizability of our framework, especially in the agent training module.
\vspace{-0.1in}
\begin{table}[!h]
    \centering
    \setlength{\tabcolsep}{1.3pt}
    \renewcommand{\arraystretch}{1.05}
    \resizebox{0.9\columnwidth}{!}{
    \begin{tabular}{c|ccc}
    \toprule[1pt]
     & \textit{Engagement} $\uparrow$ & \textit{Robustness} $\uparrow$ &  \textit{Suitability} $\uparrow$  \\ \hline
     Qwen2.5-3B-Instruct& 3.15±0.48&2.61±0.56&	52.08±0.50\\  \cdashline{1-4}[1pt/1pt]
    \ours & 3.28±0.53&2.68±0.53&	69.86±0.46  \\ 
    \bottomrule[1pt]
    \end{tabular}}
    \caption{Micro-level simulation result of \ours~on Weibo dataset.}
    \label{tab:robustness}
    \vspace{-0.4in}
\end{table}

\vpara{Comparison with existing agent architectures.} Direct comparison between our framework and traditional paradigms such as BDI is not entirely straightforward due to differing objectives and design principles. Conventional agent-based modeling approaches primarily focus on constructing realistic environments with handcrafted behavioral rules and often rely on fixed message passing mechanisms to mimic social dynamics. For example, many simulation platforms restrict information propagation to explicit follower relationships or adopt recommendation systems inspired by real-world platforms like Reddit.

% \vspace{-0.1in}
\begin{table}[!h]
    \centering
    \setlength{\tabcolsep}{1.5pt}
     \renewcommand{\arraystretch}{1.05}
    \resizebox{1.0\columnwidth}{!}{ 
    \centering
    \renewcommand{\arraystretch}{1.3}
    \begin{tabular}{c|cccc}
    \toprule[1pt]
        &Avg. Reward & Avg. Toxicity & LLM agents stances & ABM stances  \\ \hline
        \ours&\textbf{0.3661} &\textbf{0.0392} &0.5064(\textbf{0.3141}) & -\\ 
        HiSim Hybrid w/ RA & 0.1596&0.0487 &0.5340(0.1226)& 0.5067(0.1726) \\ 
        HiSim Hybrid w/ Lorenz &0.1724 &0.0471 & 0.5296(0.1187)&0.5016(0.1789) \\  \bottomrule
    \end{tabular}}
    \caption{Quantitative comparison between \ours~and HiSim-based baselines under identical settings.  \ours~achieves the highest reward and lowest toxicity, demonstrating superior intervention optimization performance.}~\label{tab:abm}
    \vspace{-0.3in}
\end{table}

In contrast, \ours~not only simulates social interactions but also leverages feedback to optimize intervention policies. For empirical comparison, we integrated existing simulation platforms into our pipeline and conducted additional experiments using HiSim~\cite{mou2024unveiling}, a hybrid LLM–ABM framework with fixed message passing. Ten-round simulations (five replications) were run under the same settings and evaluation metrics, with stance values rescaled from $[-1,1]$ to $[0,1]$ for compatibility with the ABM setup.

As shown in Table~\ref{tab:abm}, our method achieves the highest average \textbf{Reward} and the lowest text \textbf{Toxicity}, demonstrating that learned intervention policy fosters more meaningful interactions. We further analyze mean and standard deviation of \textbf{User Stances} in the final round to assess polarization and echo chamber effects (standard errors in parentheses). While the baseline environment exhibits echo chamber patterns~\cite{mou2024unveiling}, our intervention-aware model maintains a broader stance distribution, underscoring \ours’s ability to preserve opinion diversity and mitigate homogenization.

\subsection{Proofs}\label{app:proofs}
% \noindent {\textbf{Theorem 1.}  Assume the ABM model propagates a brief \( e(u_i) \in [0, 1] \) via message passing, where the social network structure is defined as a connected graph \( A \). Suppose the update mechanism for \( u_i \)'s brief takes the average of the briefs from its connected users. After propagation, each user's brief converges to be linearly inseparable.  

% This theorem reveals that linear averaging propagation in ABM inherently drives the system toward a homogeneous equilibrium, where all agents' beliefs become linearly inseparable. Consequently, diversity in agent behaviors and opinions vanishes over time, highlighting the necessity of incorporating nonlinear mechanisms or adaptive interventions to preserve heterogeneity and realistic dynamics.

\noindent\textbf{Theorem 1.}
Assume the ABM propagates a belief vector \( e(u_i) \in [0, 1] \) through message passing over a connected social network \( A \in \mathbb{R}^{n \times n} \).  Each agent \( u_i \) updates its belief by taking the average of its neighbors’ beliefs. During propagation, all agents’ beliefs converge to a homogeneous equilibrium, leading to linearly inseparable representations across agents.

\noindent {\textbf{Proof for Theorem 1.}  Since the update mechanism for \( u_i \)'s brief averages the briefs of its connected users, the propagation process can be defined as:  
\[
x^{(t+1)} = D^{-1}Ax,
\]  
where \( x \in \mathbb{R}^n \) denotes the vector of agents’ beliefs, \( D^{-1} \) denotes the inverse degree matrix, and \( D^{-1}A \) acts as a stochastic transition matrix (nonnegative entries with row sums of 1). Consequently, the network \( G \) is equivalent to a Markov chain with transition probabilities \( P = D^{-1}A \). The chain’s irreducibility and aperiodicity follow from \( G \)’s connectivity and the presence of self-loops.  

Denote the number of social simulation rounds be \( k \). For an irreducible and aperiodic Markov chain,  
\[
\lim_{k \rightarrow \infty} P^k =\lim_{k \rightarrow \infty} \{D^{-1}A\}^k =\Pi,
\]  
where \( \Pi \) is matrix with all rows equal to the stationary distribution \( \pi \), which satisfies \( \pi P = \pi \) and \( \sum_i \pi_i = 1 \). Clearly, \( \pi \) is the unique left eigenvector of \( D^{-1}A \), normalized such that all entries sum to 1. By Lemma 3.4 in~\cite{liu2020towards}, we conclude that $\Pi(x) = \frac{xD^{-1}}{\| xD^{-1}\|}$. Therefore, the update mechanism of the ABM gradually normalizes the differences between briefs, leading to reduced variance (influenced by degree) and ultimately generating indistinguishable agent roles.

\end{document}